\documentclass[12pt]{article}
\pdfoutput=1
\usepackage{epsfig}
\usepackage{amsfonts}
\usepackage{amssymb}
\usepackage{cite}
\usepackage{hyperref}
\usepackage{url}
\usepackage{subfigure}
\begin{document}

\begin{center}

\vspace{1cm}

{\bf \large SUSY  at the LHC  without Missing $P_T$} \vspace{1.5cm}

{\bf \large  A. S. Belyaev$^{1,2}$, D. I. Kazakov$^{3,4,5}$, A. Sperling$^{3}$}\vspace{0.5cm}

{\it  
$^1$School of Physics \& Astronomy, University of Southampton, UK \\
$^2$Particle Physics Department, Rutherford Appleton Laboratory, Chilton, Didcot, Oxon OX11 0QX, UK\\ 
$^3$Bogoliubov Laboratory of Theoretical Physics, Joint
Institute for Nuclear Research, Dubna, Russia.\\
$^4$Alikhanov Institute for Theoretical and Experimental Physics, Moscow, Russia\\
$^5$Moscow Institute of Physics and Technology, Dolgoprudny, Russia}
\vspace{0.5cm}

\abstract{We consider a specific class of  events of the SUSY particle production at the LHC  without missing $p_T$.  Namely, we discuss the chargino pair
production with a further decay into the W-boson and the neutralino when the masses of the chargino and neutralino differ by 80-90 GeV. In this case, in the
final state one has two Ws and missing $E_T$ but no missing $P_T$. The produced neutralinos are just boosted along Ws.  For a demonstration we consider the MSSM
with non-universal gaugino masses. In this case, such events are quite probable in the region of parameter space  where the lightest chargino and neutralino
are  mostly gauginos.  The excess in the W production cross-section  reach about 10\%
over the Standard Model background. We demonstrate that the LHC experiments, which  presently
measure  the $WW$ production cross section at the $8\%$ level can probe chargino mass  around 110 GeV within the suggested scenario, which is not accessible
via other searches. If  the precision  of $WW$ cross section measurement at the LHC will achieve the $3\%$ level, then it would probe  chargino masses up to
about 150 GeV within the no missing $P_T$ scenario.}
\end{center}


\section{Introduction}

Search for R-parity conserving Supersymmetry (SUSY) at colliders is traditionally based on the events with missing transverse momenta, 
${\not{\!P}_T}$ , which naturally appear due to the escape of the stable lightest supersymmetric particle - LSP. If the mass gap between
decaying SUSY particles (either strongly or weakly produced) and the LSP is large enough, then the LSP, usually neutralinos,  carry
considerable momenta. The triggers which are sensitive to $\not{\!P}_T$ at the level of  hundred GeV will illuminate this signature.
Unfortunately, there is no evidence for signals with $\not{\!P}_T$ so far.   There are, however, two special cases for different SUSY
signals coming from  the long-lived particles and events with no missing transverse momentum.

The first case is related to charged particles like charginos which live long enough to produce the secondary vertex or even to escape the
detector before decaying into the Standard Model (SM)  particle(s) plus neutralino. In the framework of the MSSM with the SUGRA  motivated
SUSY breaking  this might happen when the masses of the chargino and neutralino  are degenerate~\cite{GKP}.  The closer the masses the
longer is the life time. To have the secondary vertex at a few mm or more, one needs the degeneracy smaller than  few hundreds of MeV. The
gauge mediated models are more favourable for the long-lived particles, and one can easily get the particle to escape the detector
undecaying~\cite{GM}. Both scenarios, however, require the proper fine-tuning of parameters especially in the SUGRA scenario.

The second case  is less constrained. If the masses of, say, chargino and neutralino differ by 80-90 GeV, then the decay of the chargino
into  W and the neutralino is possible and goes with 100\% probability. Due to the conservation of energy in the rest frame of the
chargino, the decay products are produced almost at rest and in the laboratory frame  they are boosted along  W. As a result, the chargino
pair production gives rise to an  additional W pair production accompanied by boosted neutralinos. These neutralinos, being essentially
back to back (at the leading order) along the direction of each W-boson  carry the energy but  do not contribute to missing transverse
momentum of the event. Instead, one has the  excess of  the produced W boson pairs as compared to the Standard Model. Contrary to the
usual case, here there  will be virtually no contribution to ${\not{\!P}_T}$ from SUSY particles.

In this note, we study how probable these events are, how big the parameter space is where it happens, and what kind of excess in the W
production one might expect for a reasonable interval of chargino masses.  As an example, we consider the MSSM with non-universal gaugino
masses. We would like to note that these events, where  ${\not{\!P}_T}$ from SUSY particles does not occur at the tree level, in general
could acquire a non-negligible  ${\not{\!P}_T}$  in case of an extra hard jet radiated from the initial state quarks. In this  case, two
neutralinos will not be  balancing each other and would give a contribution to missing energy. This $W^+W^- jet+{\not{\!P}_T}$  signature,
which eventually will be suppressed in comparison with  the $W^+W^-$  one, could be still  potentially  interesting in exploring the SUSY
parameter space but is not in the scope of the current paper.

One should note that the LHC potential to probe the R-parity conserving SUSY
without using the  missing transverse momenta information has been discussed 
previously but in aspects different from the current study.
For example, in~\cite{Baer-no-missing} the
authors have shown that high lepton multiplicity events coming from the cascade decays of the
coloured SUSY partners could provide constraints on  the SUSY parameter space without using  $\not{\!P}_T$ information. In paper~\cite{Rolbiecki:2015lsa}, which is closer in spirit to our study,  the authors used the $W^+W^-$ signature to improve the limit on the light stop quarks production at the LHC. 
Finally, one should mention the case when the mass gap between the LSP and the next-to-lightest particle (NLSP) 
is small, the NLSP still decays promptly in the detector and the
pair production of the NLSP  leads to  the signature without ${\not{\!P}_T}$;
actually such a process leads to no signature  in the detector at all.
To probe such a scenario, it became traditional  now to consider  the {\it monojet signature},
when high-$P_T$ jet is radiated from the initial state  quark or gluon
and provides a monojet and ${\not{\!P}_T}$.
In our paper we study  conceptually different scenario with no  ${\not{\!P}_T}$ signature
as a key to probe SUSY at the LHC.


\section{Phenomenology of the no-missing ${\not{\!P}_T}$ MSSM scenario}

 In this section, we study  the   MSSM scenario when $m_{\chi^\pm} - m_{\chi^0}\approx M_W$
 and consider chargino production  at the LHC with a subsequent decay into W and a neutralino in the whole relevant parameter space.
 As an example  we take the MSSM  with non-universal SUSY breaking parameters~\cite{MSSM}. This does not limit our analysis since we do not rely on particular properties of the MSSM but just try to demonstrate that the advocated events are quite probable and the parameter space is not that much fine-tuned.
 
In what follows we impose the following constraint:
\begin{equation}
m_{\chi^\pm} - m_{\chi^0}\approx M_W + 0\div 10\ GeV.
\label{const}
\end{equation}
It can be easily satisfied  if both the lightest chargino and neutralino are higgsinos. Then their masses are given essentially by $\mu$ and are close. However, in this case their interaction with the light quarks in a proton is suppressed by the smallness of the Yukawa couplings, and the production cross-section is small. Therefore, we choose  the other option and consider the case when both the lightest chargino and neutralino are gauginos. In this case their masses
are defined by $M_2$ and $M_1$, respectively, and one needs the latter to be close. This practically excludes the universal scenario   where $M_2=M_1$ at the GUT scale and run down to the EW scale  reaching the ratio 2:1.
For this reason we consider the non-universal gaugino masses and treat them as free parameters at the EW scale.

All together we have the following set of parameters at the EW scale which are essential for our analysis: $M_2$, $M_1$, $\mu$ and $\tan\beta$. 
We are interested in the region of parameter space where the constraint (\ref{const}) is satisfied. Besides, we check the
chargino production cross-section in comparison with the W production in the SM and look for the regions, where
the former one gives a few percent enhancement.

\begin{figure}[ht]
\begin{center}
\leavevmode
\includegraphics[width=\textwidth]{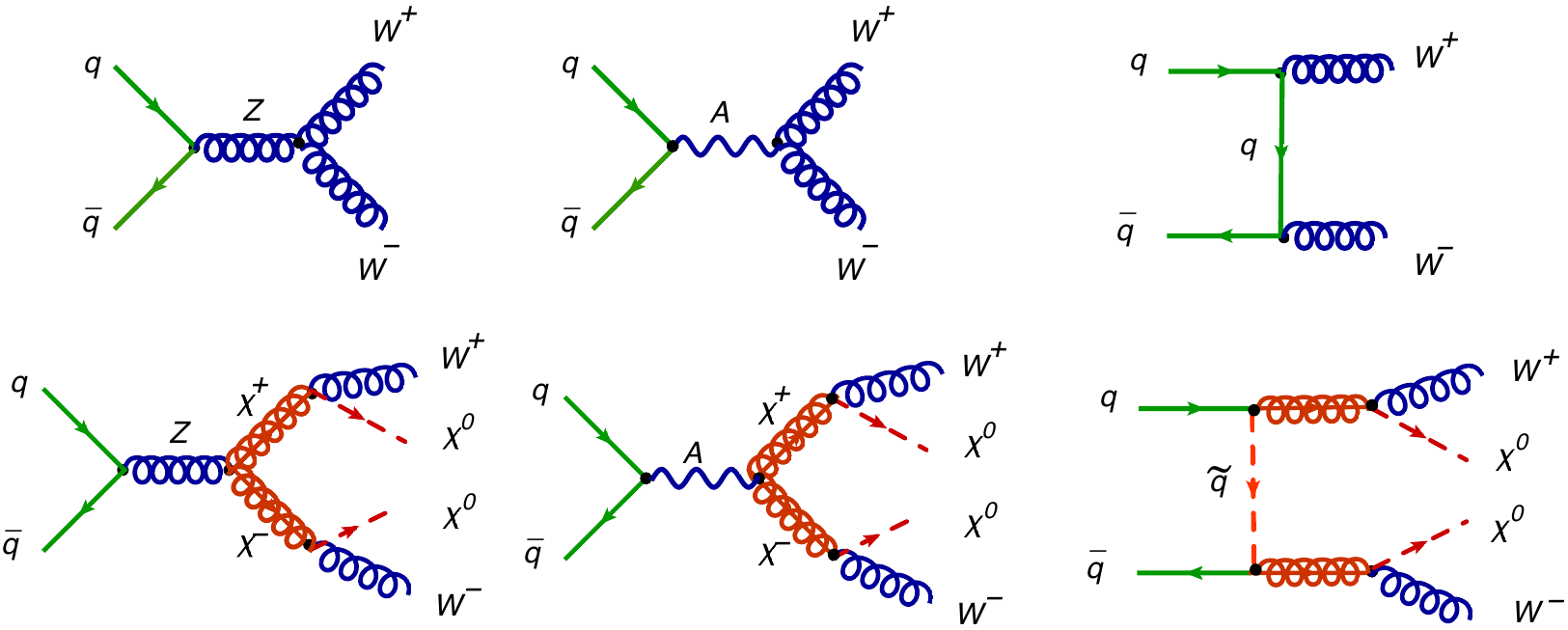}
\end{center}
\caption{The W production at the LHC: the direct one (top) and the one via the chargino decay (bottom)}
\label{process}
\end{figure}

We compare two processes of the W production: the direct SM one and the one via chargino decay.
The corresponding Feynman diagrams  are  shown in Fig.\ref{process}.
The leading order (LO) SM cross-section for the W pair production at the LHC with the cm energy of 8 TeV and 13 TeV is equal to 35.7 and 67.7 pb, respectively.  
To evaluate  it together with the   $\chi^+\chi^-$  production, we use the 
CalcHEP 3.6.15 code~\cite{CalcHEP} with  the parton distributions MRST2008lo68cl~\cite{pd} 
linked to CalcHEP via LHAPDF6~\cite{LHAPDF} framework. For this evaluation, the  renormalisation  and factorisation
scales were set to $\mu_R = \mu_F = M_W$ and the EW parameters were set to reproduce  $G_F = 1.16639 \times  10^{-5}$~GeV$^{-2}$.
Our LO results for the $WW$ production agree  with those from~\cite{WW-NNLO},
where the authors went up to the NNLO level.
They found out that at the LHC energies  the  NLO order K-factor for the W production is about 1.5,
while the NLO K-factor for chargino pair production for chargino masses of the order of 100 GeV is about 1.35~\cite{chi+chi-NLO} .
Hence, at the NLO the ratio of $\sigma_{\chi^+\chi^-}/\sigma_{WW}$ will be slightly reduced by a factor of $1.35/1.5=0.9$
as compared to the tree level result. We take it into account in our estimations:
\begin{equation}
\frac{\sigma_{\chi^+\chi^-}}{\sigma_{WW}}
\equiv
\frac{\sigma^{NLO}_{\chi^+\chi^-}}{\sigma^{NLO}_{WW}}=\frac{\sigma^{LO}_{\chi^+\chi^-}}{\sigma^{LO}_{WW}} \times \frac{K^{NLO}_{\chi^+\chi^-}}{K^{NLO}_{WW}} \simeq \frac{\sigma^{LO}_{\chi^+\chi^-}}{\sigma^{LO}_{WW}}\times 0.9
\end{equation}

For our analysis we use the MSSM model implemented into CalcHEP which is publicly available at High Energy Physics Model Database (HEPMDB)
at \url{http://hepmdb.soton.ac.uk/hepmdb:0611.0028}.
One should note that the present theoretical accuracy of $pp\to W^+W^-$ production at the LHC 
as defined by its NNLO results (59.84pb$^{+2.2\%}_{-1.9\%}$\cite{WW-NNLO})
is at about $2\%$, 
while ATLAS ($71.4\pm 5.6$~pb\cite{WW-ATLAS}) and CMS ($60.1\pm 4.8$~pb\cite{WW-CMS})
currently measure  this process with the accuracy of about 8\%. One can also see from the numbers given above that ATLAS 
observes about $2\sigma$ excess in $WW$ production. Eventually, this excess is not conclusive;
however, it provides us with some source of speculation. In particular, we would like to verify if
MSSM could provide enough  contribution from $\chi_1^+\chi^-$ with no ${\not{\!P}_T}$ kinematics to explain this excess quantitatively.

To calculate the cross-section in the case of chargino decay, we actually calculate the chargino production cross-section and assume that the chargino is produced on shell and subsequently decays into W and the neutralino provided $m_{\chi^\pm}\geq m_W+m_{\chi^0}$. 
In doing so we perform the scan of the whole relevant parameter space mentioned above:
$M_1[-500-500\mbox{~GeV}]$, $M_2[10-500\mbox{~GeV}]$, $\mu[10-2000\mbox{~GeV}]$ and $\tan\beta[10-50]$,
imposing the constraint (\ref{const}) and the  LEP2  limits on chargino mass reaching  103.5 GeV.
To evaluate the cross sections and SUSY mass spectrum, we use the standard
CalcHEP subroutines.
One should note that in the
region (\ref{const}) the  LHC  cannot set better limits than LEP2 in the generic scenario
when sleptons, squarks and gluino are heavy because of very low ${\not{\!P}_T}$ (see e.g.~\cite{CMS-chargino}).
\begin{figure}[htb]
\hspace*{-0.3cm}\includegraphics[width=0.53\textwidth]{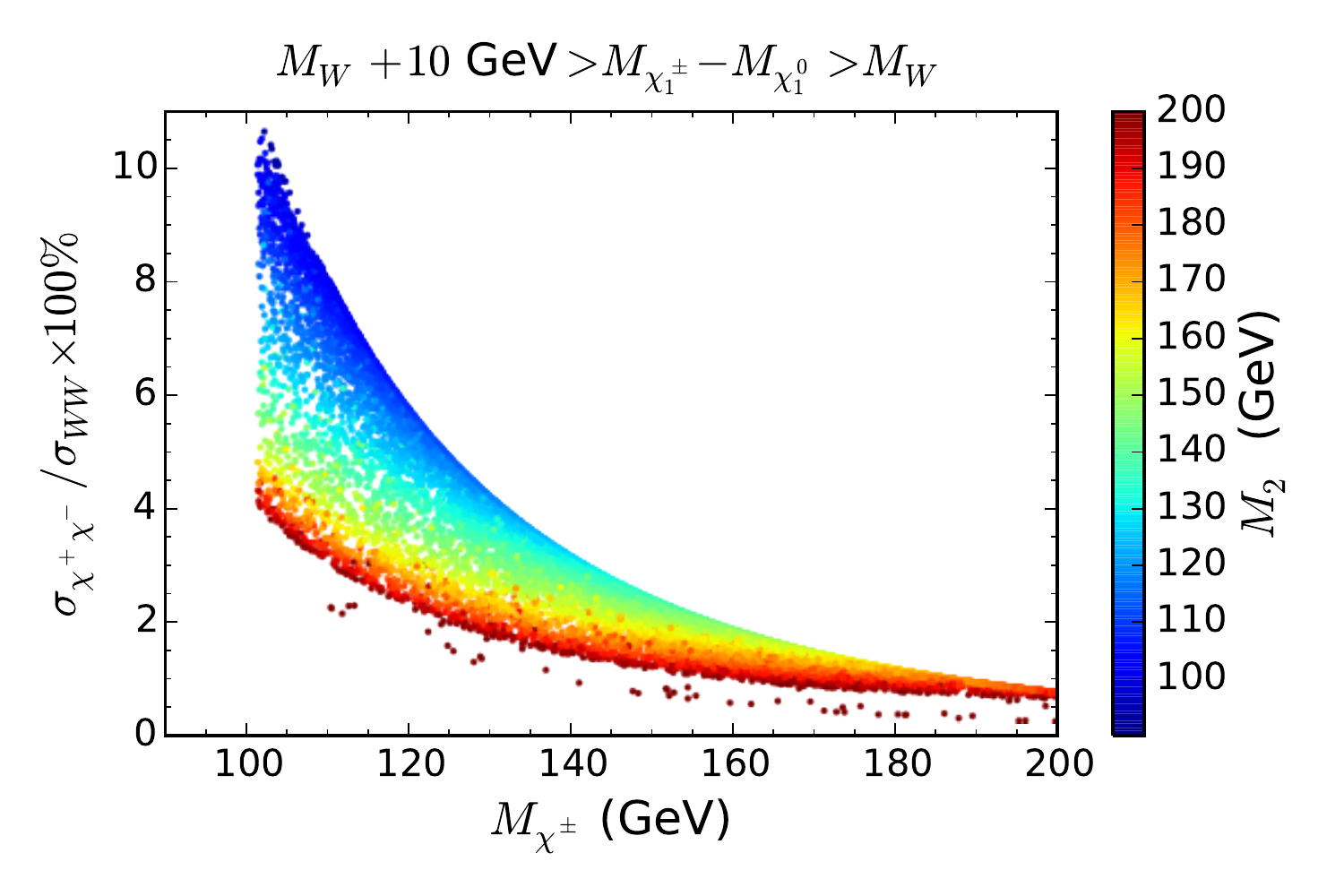}%
\hspace*{-0.3cm}\includegraphics[width=0.53\textwidth]{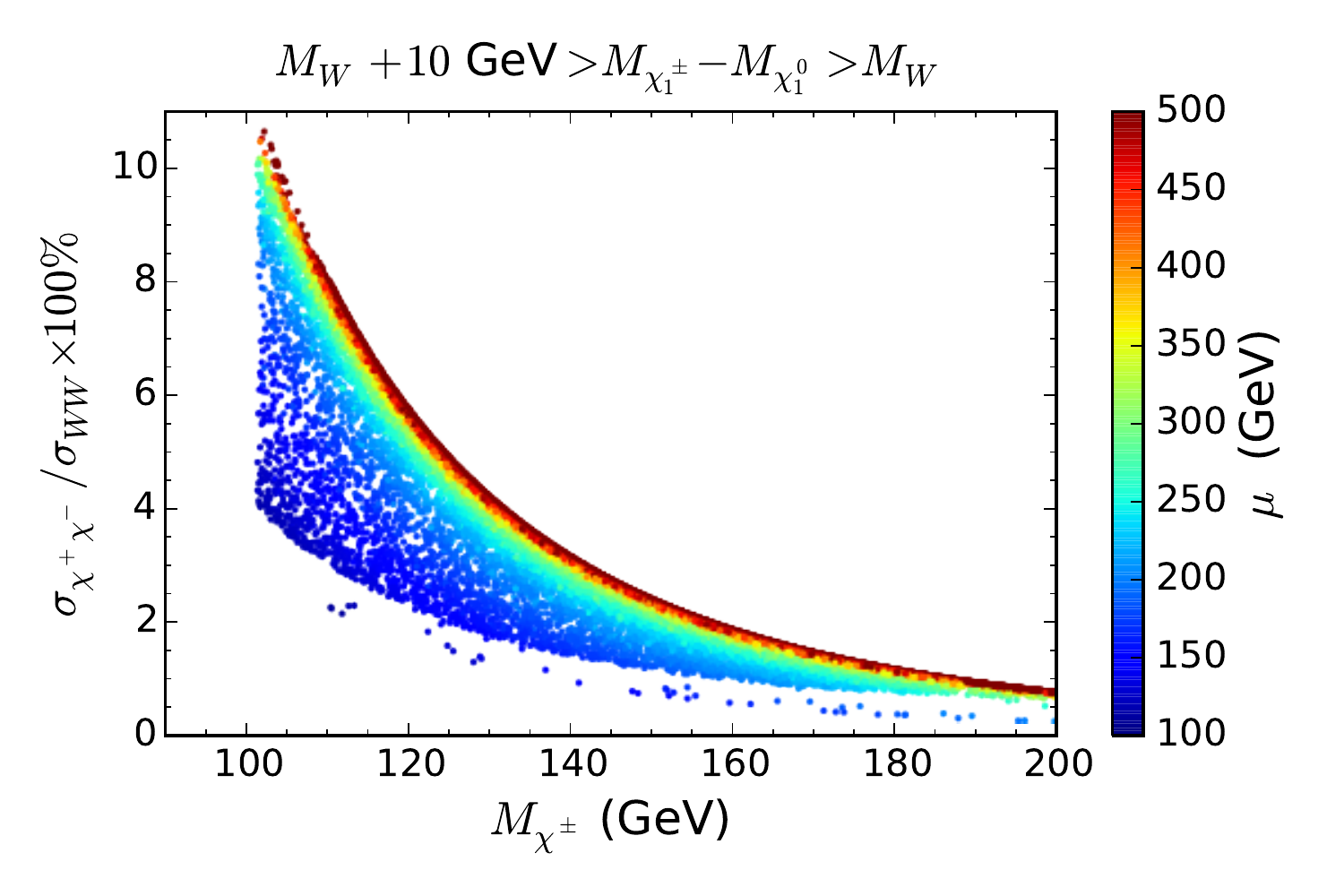}
\caption{\label{csr}
$\frac{\sigma_{\chi^+\chi^-}}{\sigma_{WW}}$ at the LHC@8TeV versus chargino mass
as a colour map of $M_1$ (left) and  $\mu$ (right).}
\end{figure}

The results of the scan are shown in Fig.~\ref{csr},
where we present the ratio $\frac{\sigma_{\chi^+\chi^-}}{\sigma_{WW}}$ at the LHC@8TeV versus chargino mass
as a colour map of $M_1$ (left) and  $\mu$ (right).
One can see that the chargino can  contribute up to about 10\% to the $W^+W^-$ production at the LHC 
if chargino mass is of the order of 100 GeV and just above the LEP2 constraints.
One can also see from this figure that $\sigma_{\chi^+\chi^-}$ achieves its highest value (for a given chargino mass)
for minimal $M_2$ around 100 GeV and high values of $\mu$ parameter, $\mu\gtrsim 400$~GeV.
In this parameter space chargino has a large wino component and, respectively, large coupling to the Z-boson,
the main mediator of the chargino pair production. The largest wino component of the chargino
defines the upper band of its production  cross section.
 On the contrary, when  $\mu$ is small and  $M_2$ is large, the chargino has a dominant higgsino component, suppressed coupling to Z-boson
and, respectively, a low production cross section indicated by the lower band in Fig.~\ref{csr}.

\begin{figure}[htb]
\centerline{
\includegraphics[width=0.7\textwidth]{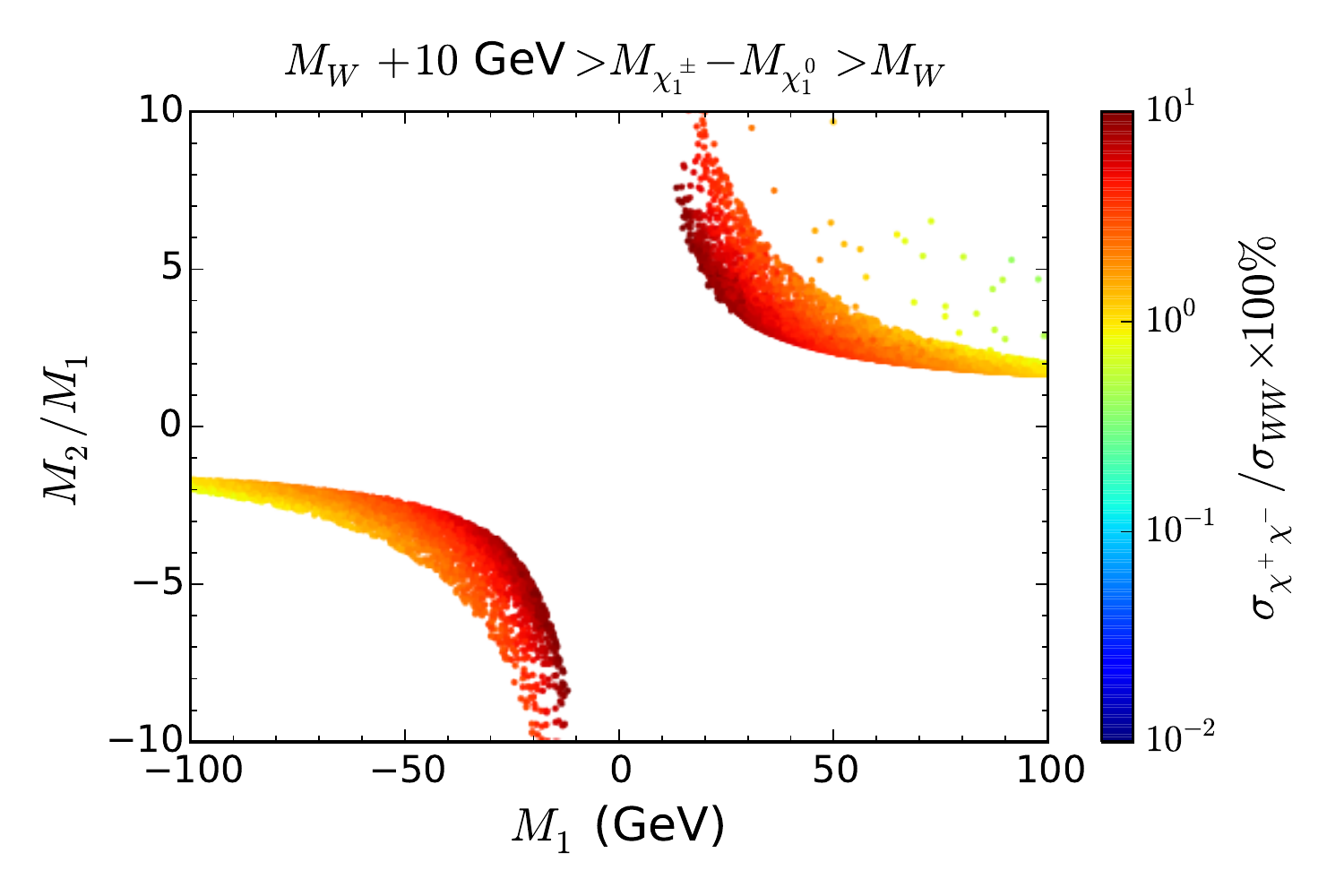}
}
\caption{\label{M2M1r}
The ratio $M_2/M_1$ versus $M_1$ with the colour map of $\frac{\sigma_{\chi^+\chi^-}}{\sigma_{WW}}$, assuming the constraint (\ref{const}) is satisfied.}
\end{figure}

In Fig.~\ref{M2M1r}, we present  the ratio $M_2/M_1$ versus $M_1$ as a scatter  plot from our scan
with the colour map indicating  the ratio $\frac{\sigma_{\chi^+\chi^-}}{\sigma_{WW}}$. One  can see from this plot that
the value of the ratio $\frac{\sigma_{\chi^+\chi^-}}{\sigma_{WW}}$ above a few percent is achievable only for $M_2/M_1>3$
and reaches the maximum for $M_2/M_1\simeq 5-6$
where $m_{\chi^+}$ is just above the LEP2 constraints and $m_{\chi^0}$ is about 20 GeV.
It is clear that such a scenario cannot be realized for universal gaugino masses at the GUT scale
which gives  $M_2/M_1\simeq 2$ at the electroweak scale.

\begin{figure}[htb]
\centering
\includegraphics[width=0.5\textwidth]{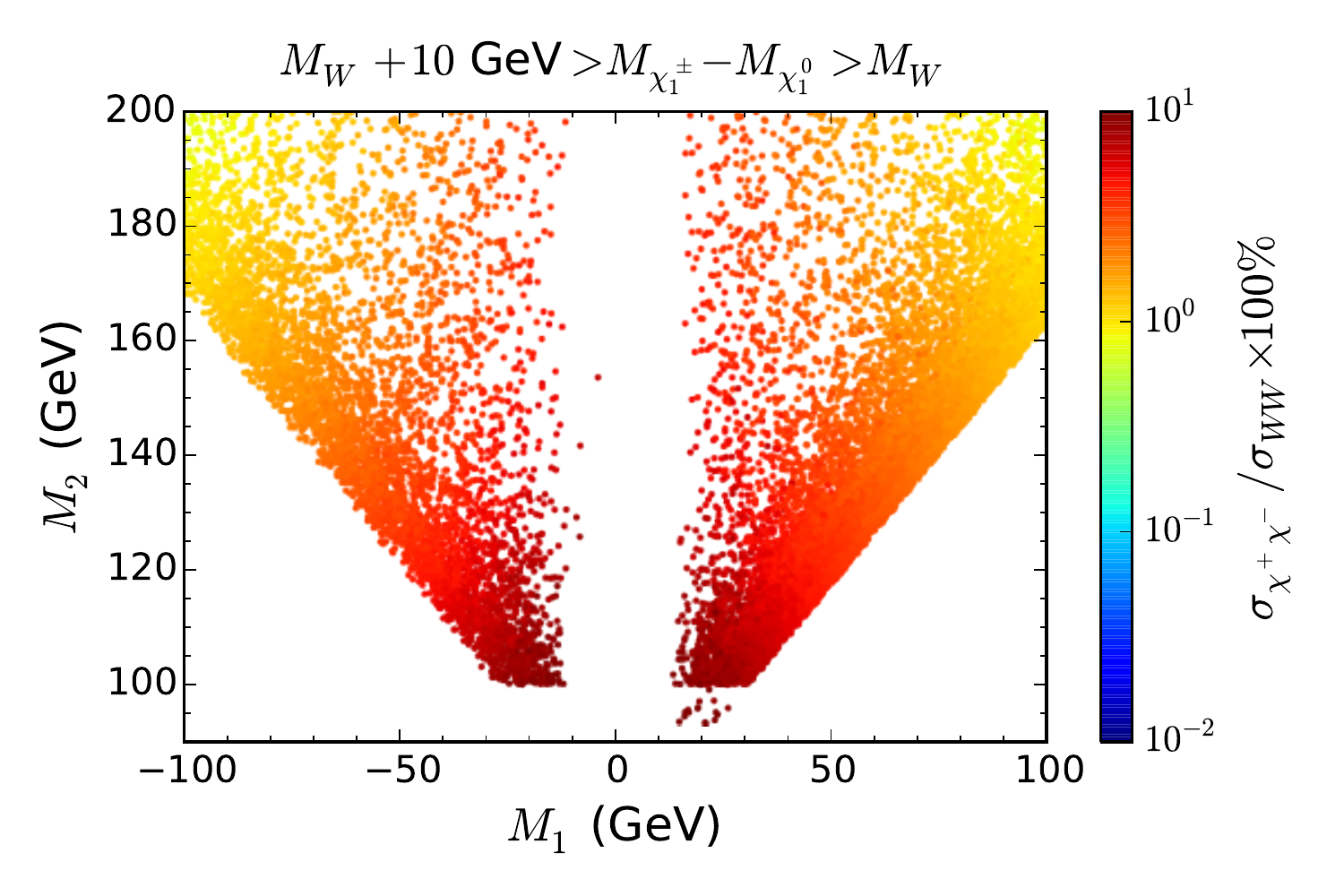}%
\includegraphics[width=0.5\textwidth]{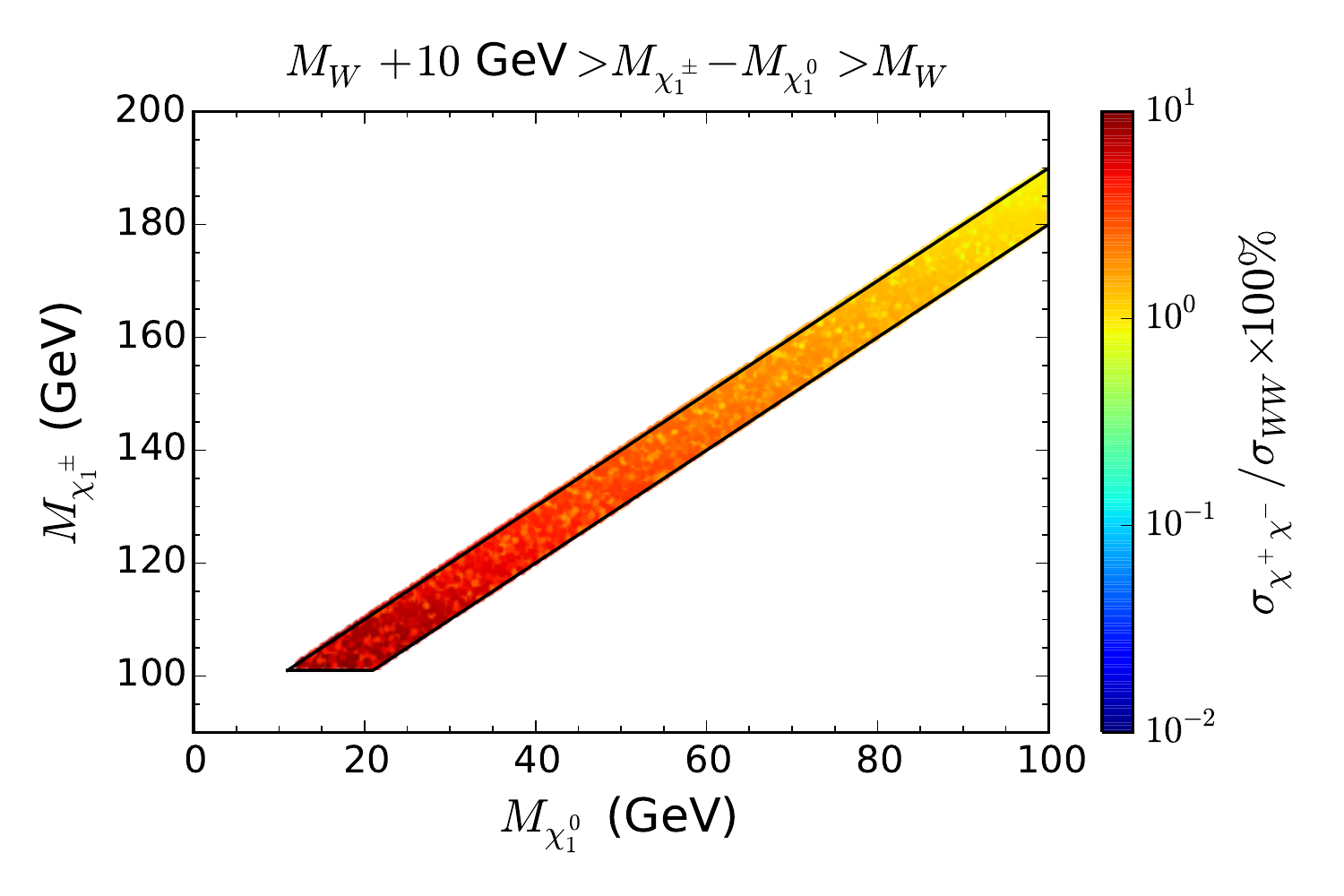}\\
\includegraphics[width=0.5\textwidth]{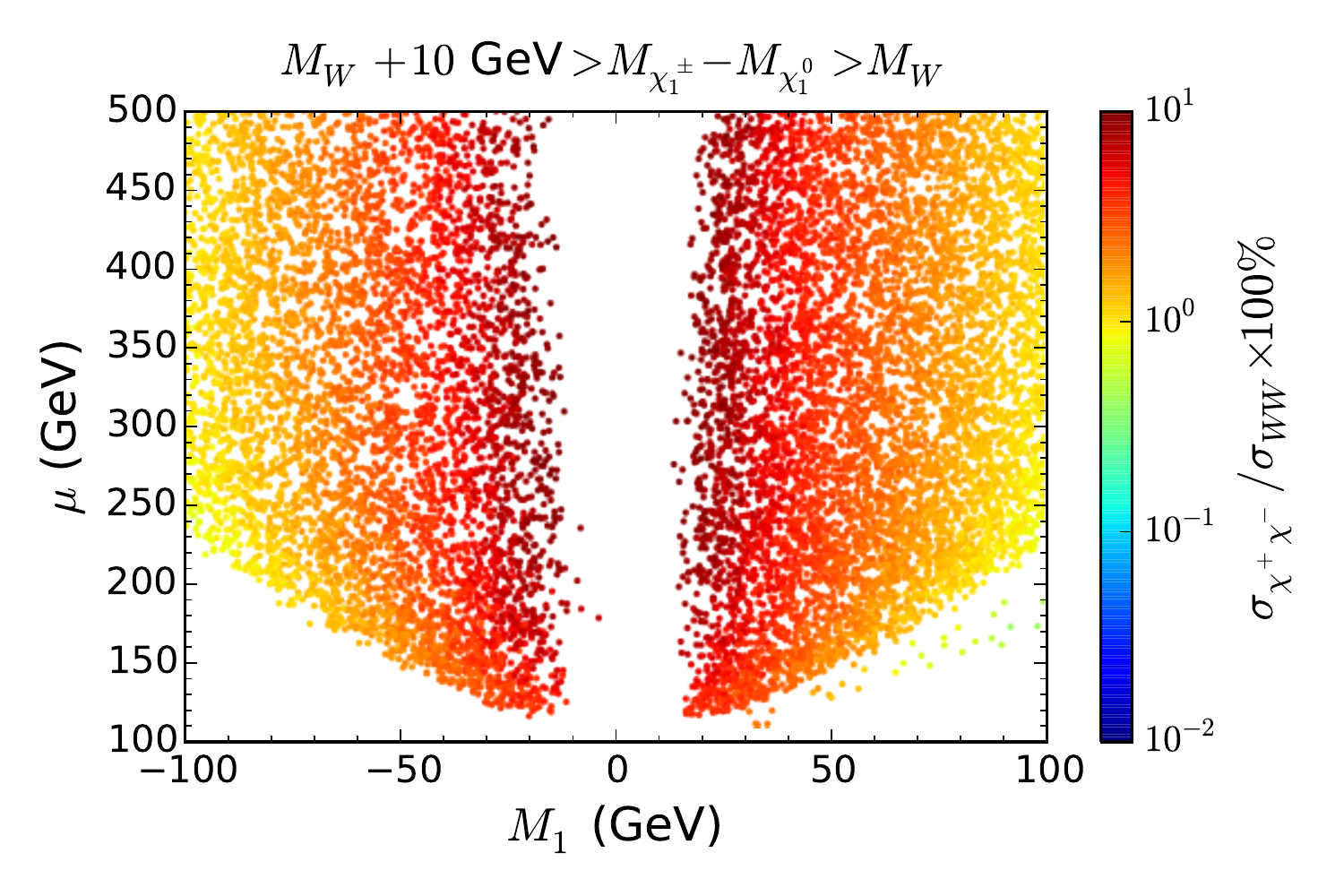}%
\includegraphics[width=0.5\textwidth]{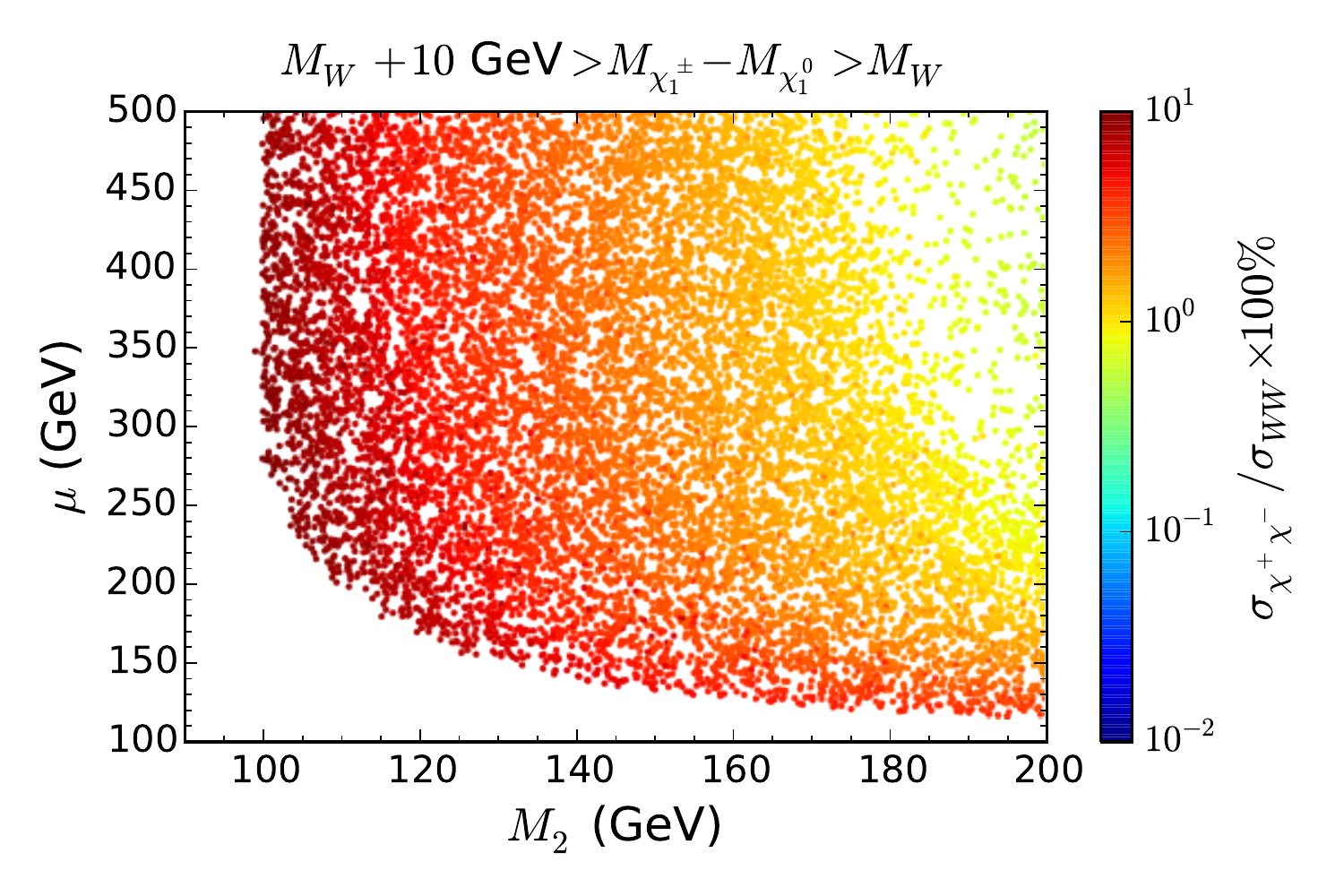}\\
\includegraphics[width=0.5\textwidth]{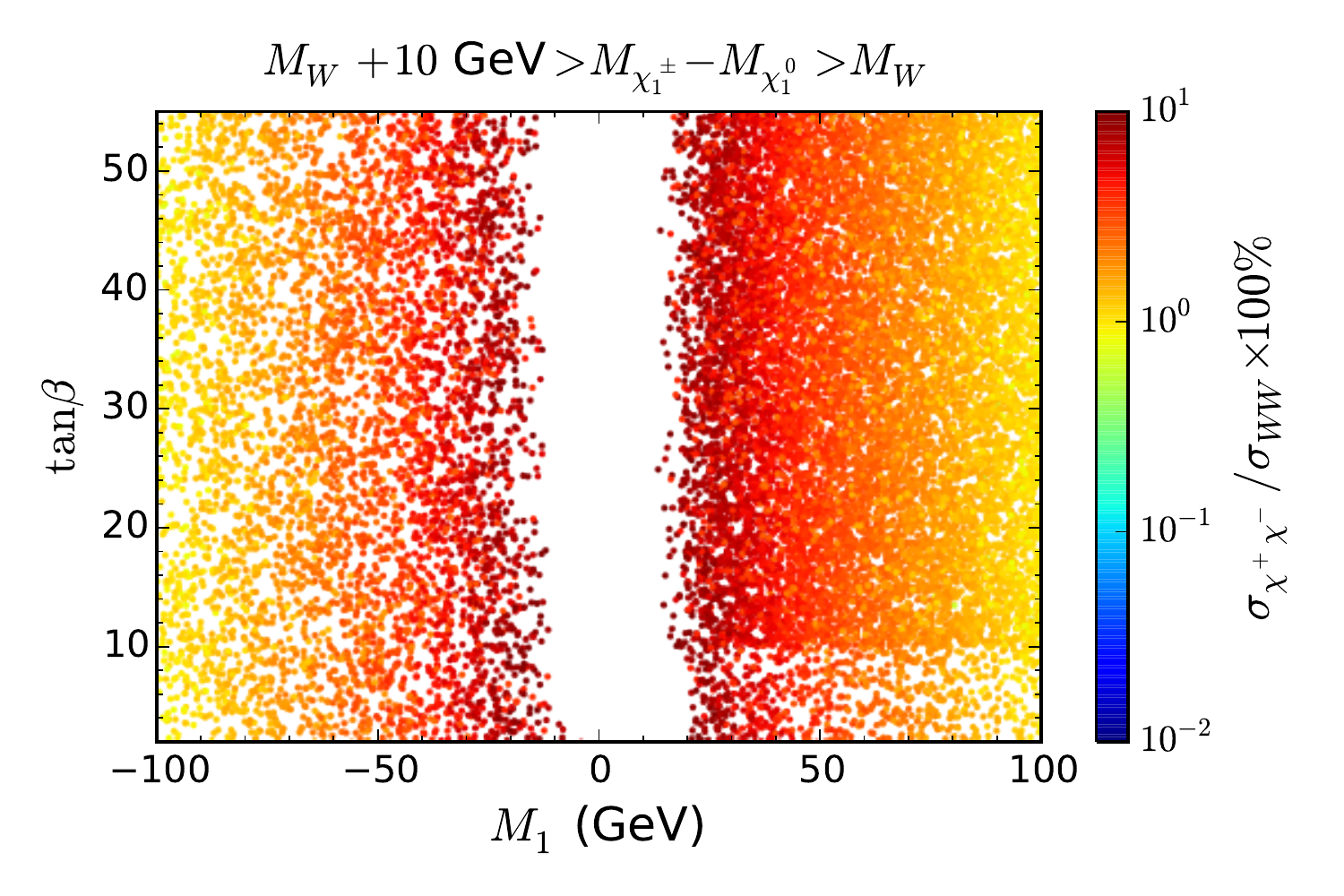}%
\includegraphics[width=0.5\textwidth]{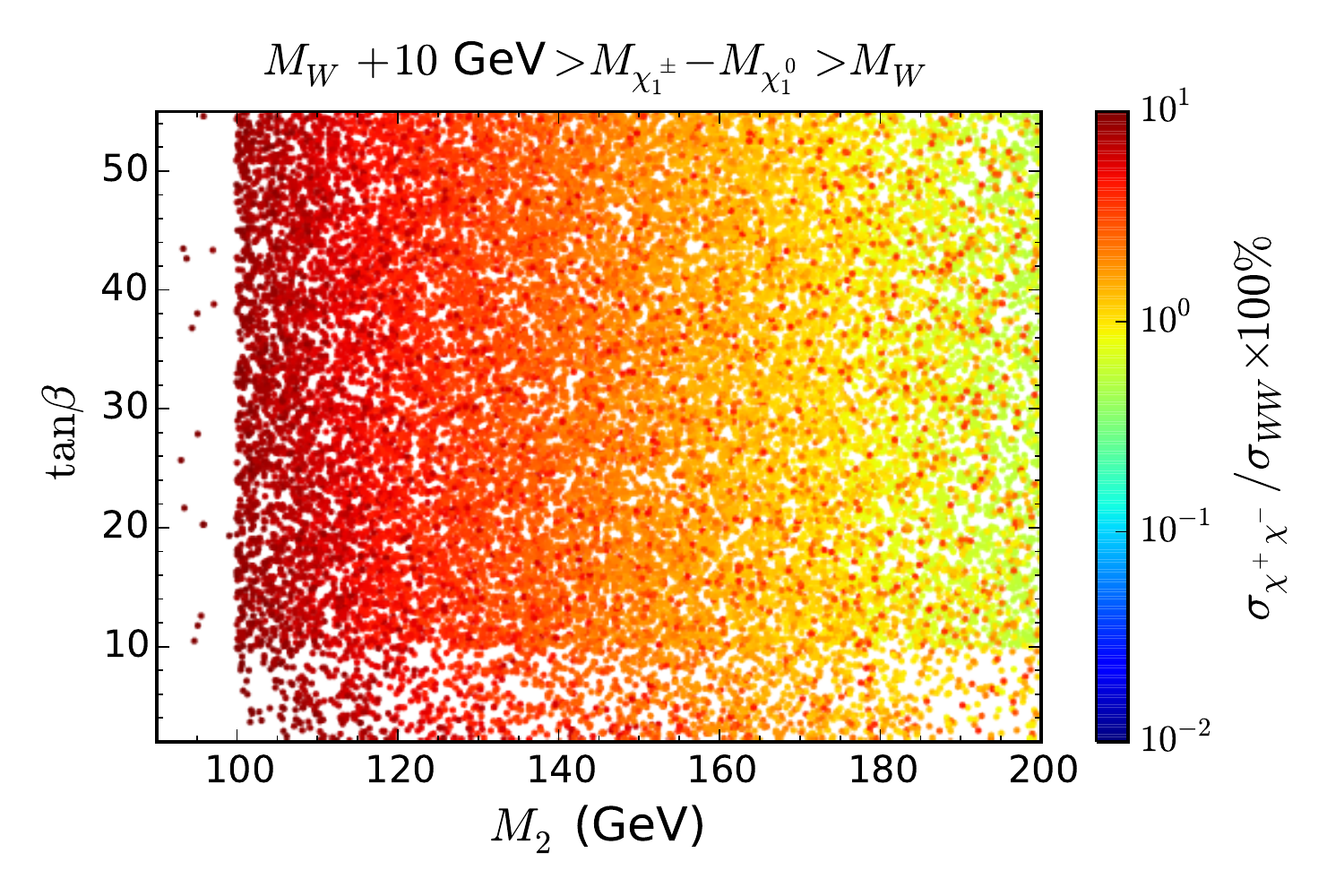}
\caption{The  2-dimensional scatter plots for
various pairs of the model parameters and physical masses from the scan of the low energy MSSM parameter space assuming that constraint (\ref{const}) is satisfied.
The colour map indicates the ratio of the cross-sections of chargino production in the MSSM to the cross-section of W production in the SM at LHC@8TeV.}
\label{region}
\end{figure}

It is also very informative to look at the other 2-dimensional scatter plots for
various pairs of the model parameters and physical masses
presented in Fig.\ref{region}.
One can see that practically all the values of the parameters are allowed including the values of $\mu$ and $\tan\beta$;
however, if one wants to increase the cross-section one is pushed to the edges of parameter space towards smaller values. These regions are favoured by small chargino and neutralino masses since the cross section of chargino production is inversely proportional to the latter. Note, however, that the ratio of $M_{\chi^+}/M_{\chi^0}$ is constrained to the narrow band where their mass difference satisfies the constraint (\ref{const}). This band is linear but does not give the fixed mass ratio since it does not go through the origin of the coordinate plane. For small masses one finds that the neutralino mass is very small, which is still not excluded by modern data.

\begin{table}[htb]
\begin{center}
\begin{tabular}{|| l || l | l |l| l || l |l || l| l ||}
\hline
\hline
BM$\#$ & $\tan\beta$ &  $M_1$ & $M_2$ & $\mu$ &  $m_{\chi^0}$ &  $m_{\chi^+}$ &  $\sigma_{\chi^+\chi^-}^{8TeV} (pb) $ & $\sigma_{\chi^+\chi^-}^{13TeV} (pb) $\\
\hline
\hline
1&	20&	     20&  100&   400  &  19.02 &	  105.1 &	    3.73 &   7.19\\
\hline
2&	40&	     22&  100&   1000 &  21.65 &	  110.0 &	    3.23 &   6.22\\
\hline
3&	10&	     26&  105&   1400 &  25.33 &	  115.0 &	    2.74 &   5.32\\
\hline
4&	10&	     32&  110&   1000 &  31.12 &	  119.6 &	    2.34 &   4.59\\
\hline
5&	20&	     40&  115&   800  &  39.23 &	  125.2 &	    1.96 &   3.87\\
\hline
6&	10&	     44&  120&   1000 &  43.00 &	  130.4 &	    1.68 &   3.35\\
\hline
7&	10&	     48&  125&   800  &  46.80 &	  135.0 &	    1.47 &   2.94\\
\hline
8&	20&	     54&  130&   600  &  52.85 &	  140.0 &	    1.27 &   2.55\\
\hline
\hline
\end{tabular} 
\end{center}
\caption{The cross-section of the chargino production at various benchmark points  at the LHC at  8 and 13 TeV. 
The masses of the chargino and the neutralino as well as values for $M_1$, $M_2$ and $\mu$ parameters are given in GeV}\label{tab:sigma}
\end{table}
\begin{figure}[htb]
\centerline{
\includegraphics[width=0.70\textwidth]{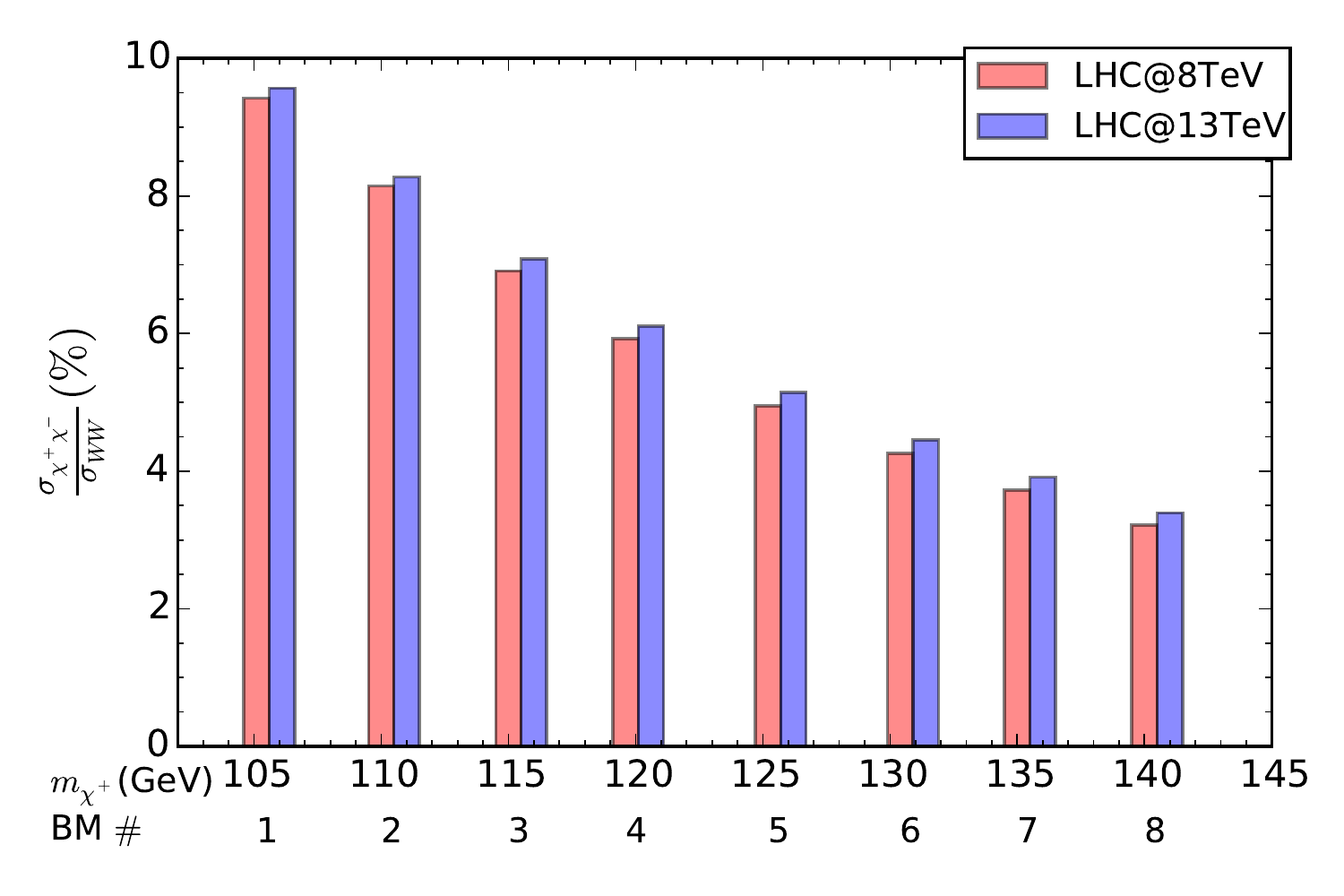}
}
\caption{The ratio of the W production cross-section via the chargino decay to the SM cross-section at the LHC for the cm energy of 8 and 13 TeV for various benchmark points from Table~\ref{tab:sigma}.\label{fig:sigma}}
\end{figure}

There is no real preference for any particular values of $\tan\beta$ coming from constraint (\ref{const}). This preference will appear if one considers the other constraints, in particular, the amount of the Dark matter~\cite{deB}. The $\mu$ parameter might also vary in a wide
range but is typically bigger than $M_1$ and even $M_2$ that provides the gaugino origin of the lightest chargino and neutralino.
Thus, one can see that the fulfillment of the requirement (\ref{const}) is not that restrictive and does not require much fine-tuning. One has at least 4 free parameters which can be used to get into the desired region. Even more possibilities appear in extended models. We use here the  MSSM as an illustration of the most constrained model. Of course, if one wants to apply more constraints, they might get into  contradiction with this one. For instance, the LHC limits already excluded a large part of the parameter space. However, on the other hand, the LHC limits are based on the  events with ${\not{\!P}_T}$  and might miss some opportunities. One can think of a model free case and just look for advocated events.
The chargino production cross-section in its turn depends on its mass and the mixings. As an illustration, we calculate it for several benchmark points 
shown in Table~\ref{tab:sigma} for the LHC energies of 8 and 13 TeV.
 The ratio $\frac{\sigma_{\chi^+\chi^-}}{\sigma_{WW}}$ for the benchmark points from Table~\ref{tab:sigma}  is visualised
  in Fig.~\ref{fig:sigma}, where one can see that the W production cross-section via the chargino decay varies 
 between 9 and 4\% for the chargino mass range between 105 and 140 GeV. The cross-section   ${\sigma_{\chi^+\chi^-}}$
 is close to its maximum when $\mu>M_2$. 
 In this case the $\frac{\sigma_{\chi^+\chi^-}}{\sigma_{WW}}$ is defined mainly by the chargino mass
 and its decrease is caused mainly by the parton distributions.  
 Thus, taking higher masses one gets a smaller cross-section and  hence a smaller excess in the W production.  

We would like to stress once more  that our calculations within the  MSSM serve as an illustration of the possibility of  the SUSY production process without missing $P_T$. Precise values of the cross section and the position of the benchmark points in parameter space are not essential. In a more complicated model these numbers may change but the very possibility of the advocated process remains.

\section{Discussion}

We have explored the contribution of SUSY particles to the $WW$ production with no ${\not{\!P}_T}$  signature.
We demonstrate that the LHC experiments which  presently measure  the $WW$ production cross section at the $8\%$ level
can probe chargino masses  around 110 GeV within the suggested scenario, which is not accessible via other searches.
If  the precision  of the $WW$ cross-section measurement at the LHC will achieve $3\%$ level, then it would probe  the chargino masses up to about 150 GeV within the no missing $P_T$ scenario.

This  excess of Ws produced via the chargino decay might be noticeable.  The recent data on the diboson production with a subsequent decay into muons  has
shown that all results are compatible with the theoretical expectation within the statistical and systematic uncertainties though some excess with respect to
the SM expectations at the level of two $\sigma$  is observed by the ATLAS collaboration~\cite{WW-ATLAS}. This might be the usual fluctuation but equally might
indicate the manifestation of a new physics. The detailed analysis of these data as well as the new run results will clear the case. However, this kind of
processes is precisely the one where one can expect the new physics to show up.  Possible SUSY interpretation of this excess was considered in a sequence of
papers~\cite{Curtin} and it was shown that the data from both experiments can be better fitted with the inclusion of electroweak gauginos with masses of O(100)
GeV. We have demonstrated that in the case of low energy supersymmetry,  these processes are quite natural.  The mass range of  the electroweak gauginos might
be even higher being well within the limits of the LHC searches. They might come with as well as without missing $p_T$. The latter possibility, which is the
main concern of our paper, is less probable but is quite possible and should be taken into account in analysis.

\section*{Acknowledgements}
DK and AS acknowledge financial support from the RFBR grant \# 14-02-00494  and the Heisenberg-Landau Program, Grant \# HLP-2014-07.
AB acknowledges partial support from the STFC grant ST/L000296/1,
the NExT Institute , Royal Society Leverhulme Trust Senior Research Fellowship LT140094 and
Soton-FAPESP grant.

\end{document}